\documentclass[pre,showkeys,english,showpacs,twocolumn,aps,prb,a4paper,floatfix]{revtex4}
\usepackage[T1]{fontenc}
\usepackage{color}
\usepackage[latin1]{inputenc}
\usepackage{graphicx}
\usepackage{bm}
\usepackage{epsfig}
\usepackage{amsmath}
\usepackage{amssymb}
\usepackage{amssymb}

\begin{document}

\title{Uniform phases in fluids of hard isosceles triangles: one component and binary mixtures}

\author{Yuri Mart\'{\i}nez-Rat\'on}
\email{yuri@math.uc3m.es}
\author{Ariel D\'{\i}az-De Armas}
\email{ardiaza@math.uc3m.es}
\affiliation{Grupo Interdisciplinar de Sistemas Complejos (GISC), Departamento 
de Matem\'aticas, Escuela Polit\'ecnica Superior, Universidad Carlos III de Madrid, 
Avenida de la Universidad 30, E-28911, Legan\'es, Madrid, Spain}
\author{Enrique Velasco}
\email{enrique.velasco@uam.es}
\affiliation{Departamento de F\'{\i}sica Te\'orica de la Materia Condensada, Instituto de 
F\'{\i}sica de la Materia Condensada (IFIMAC) and Instituto de Ciencia de Materiales 
Nicol\'as Cabrera, Universidad Aut\'onoma de Madrid, E-28049, Spain}

\begin{abstract}
We formulate the scaled particle theory for a general mixture of hard isosceles triangles and 
calculate different phase diagrams for the one-component fluid and for certain binary mixtures. The fluid 
of hard triangles exhibits a complex phase behavior: (i) the presence of a triatic phase 
with sixfold symmetry, (ii) the isotropic-uniaxial nematic transition is of first order 
for certain ranges of aspect ratios, and (iii) the one-component system exhibits nematic-nematic 
transitions ending in critical points. We found the triatic phase to be stable not only for equilateral 
triangles but also for triangles of similar aspect ratios. We focus the 
study of binary mixtures on the case of symmetric mixtures: equal particle areas with
aspect ratios ($\kappa_i$) symmetric with respect to the equilateral one: $\kappa_1\kappa_2=3$. 
For these mixtures we found, aside from first-order isotropic-nematic and nematic-nematic
transitions (the latter 
ending in a critical point): (i) A region of triatic phase stability even for mixtures made of 
particles that do not form this phase at the one-component limit, and (ii) the presence of a Landau 
point at which two triatic-nematic first-order transitions and a nematic-nematic 
demixing transition coalesce. 
This phase behavior is analog to that of a symmetric three-dimensional mixture
of rods and plates.
\end{abstract}

\date{\today}



\maketitle

\section{Introduction}

Two dimensional fluids of hard anisotropic particles are paradigms 
of systems where liquid-crystal (LC) phases can be stabilized solely by entropy.
Hard-rod particles  
such as hard rectangles (HR), discorectangles (HDR) or ellipses (HE), exhibit the
completely disordered isotropic (I) phase, but also a nematic (N) phase at 
higher densities where particle axes point, on average, along a common director. In two dimensions (2D) the N phase does not
possess true long-range orientational order, and the 
N-I transition is usually continuous via a Kosterlitz-Thouless disclination unbinding 
mechanism. The N phase is stable for high enough aspect ratios and 
its stability region (in the density-aspect ratio phase diagram) 
is bounded below by the I phase, and above by other LC nonuniform phases such as the 
smectic (S) or completely  ordered crystal (K) phases. 
 At low aspect ratios the I phase can exhibit 
a direct transition to a plastic crystal 
(PK) or to a more complex crystalline phase in which particle shapes, orientations
and lattice structures are coupled in a complex fashion. The phase behavior
of HDR was studied in detail by MC simulations \cite{Bates,Wittmann} 
and theory \cite{MR1,Ariel,Wittmann}. 
This particle shape, as well as the elliptical one 
\cite{Cuesta1,Bautista,Schlacken}, can stabilize the I, N, PK, and K 
phases with HDR exhibiting a region of S stability at high densities. 
However a fluid of HR with sufficiently small aspect ratio can also stabilize a tetratic (T) phase 
\cite{MR1,Schlacken,MR2,Donev,Triplett,Anderson,Escobedo,Lowen} with fourfold symmetry: 
the orientational 
distribution function is invariant under $\pi/2$ rotations. This peculiar 
liquid-crystal texture was also found in experiments on 
colloidal \cite{Chaikin} and non-equilibrium granular \cite{Narayan,Dani1,Miguel1,Walsh} 
systems. Particles with a more complex shape, such as zigzag particles, or hockey stick-shaped 
particles, exhibit interesting phase behaviors, such as the increase of  
S stability with respect 
to the N phase, by changing the particle shape. The former, having quasi-ideal layers,   
can be made of particles tilted with respect to the layer normal, and it can also be 
antiferromorphic \cite{Szabi1,Charbonneau,Szabi2}.

The effect of confinement on the phase behavior of two-dimensional or quasi two-dimensional
hard-rod fluids has 
also been extensively studied \cite{Daniel2,Daniel3,Mulder,Miguel2,Teixeira}. The symmetry 
of the confining external potential can (i) 
change the relative stability between 
different LC phases with respect to the bulk and (ii) 
induce the presence of topological defects in the N director field.
On the other hand, mixtures of anisotropic particles in 2D can exhibit, in analogy with 
their 3D counterparts, 
different demixing scenarios \cite{yuri_demix,dani_demix,yuri_ellipses}. However, 
I-I demixing was not found in mixtures of two-dimensional hard bodies \cite{Talbot}. 

The aim of the present article is to develop the scaled-particle theory (SPT), a
second-virial-based theory, for freely-rotating hard-triangle (HT) 
mixtures, to study 
the phase behavior of the one-component fluid and of certain binary mixtures. 
Our effort is focused on elucidating the effect of two-body interactions 
on the symmetry of orientationally ordered phases. Previous experimental, 
theoretical, 
and simulation works on equilateral \cite{Zhao,Scott,Benedict,Dijkstra} 
and right-angled \cite{Dijkstra} HT predicted 
the presence of triatic (TR) \cite{Zhao,Dijkstra} and rhombic (R) \cite{Dijkstra} 
nematic phases, respectively, with orientational distribution functions $f(\phi)$ having six-fold 
[$f(\phi)=f(\phi+\pi/3)$] or eight-fold [$f(\phi)=f(\phi+\pi/4)$] symmetries.
The controlled synthesis
of dispersed colloidal systems, made by microlithography, constitutes an experimental
realization of fluids of hard Brownian particles of precisely controlled shapes. The roughness
controlled depletion attractions allowed and experimental exploration of dense two-dimensional
systems of hard equilateral triangles \cite{Zhao}, which are one system of hard achiral
particles exhibiting local \cite{Zhao} or long-ranged \cite{Dijkstra} chiral ordering, 
with the most prominent chirality observed in lattice structures near close packing \cite{Dijkstra}.

\begin{figure}
	\epsfig{file=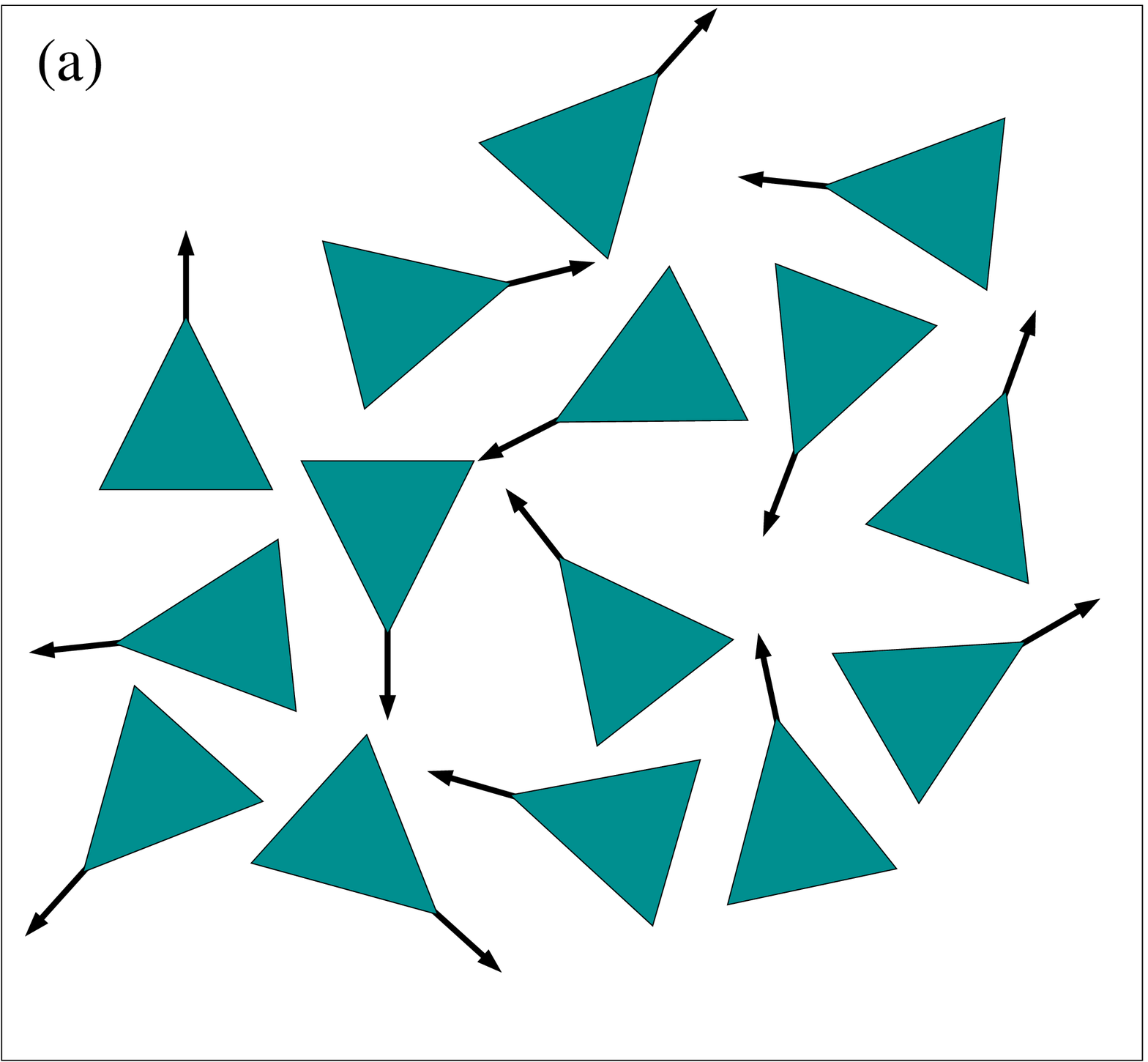,width=1.1in}
	\epsfig{file=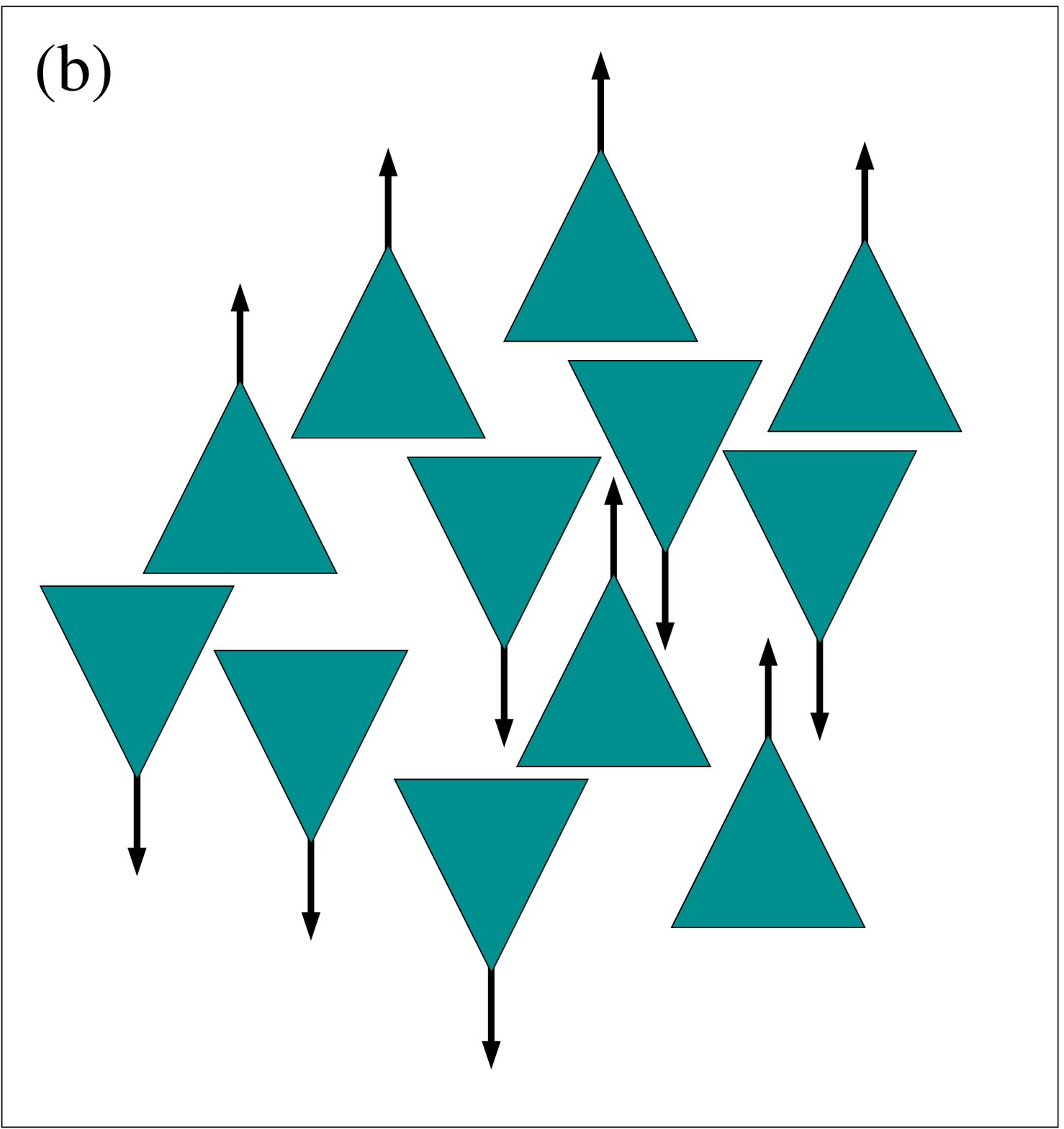,width=0.95in}
	\epsfig{file=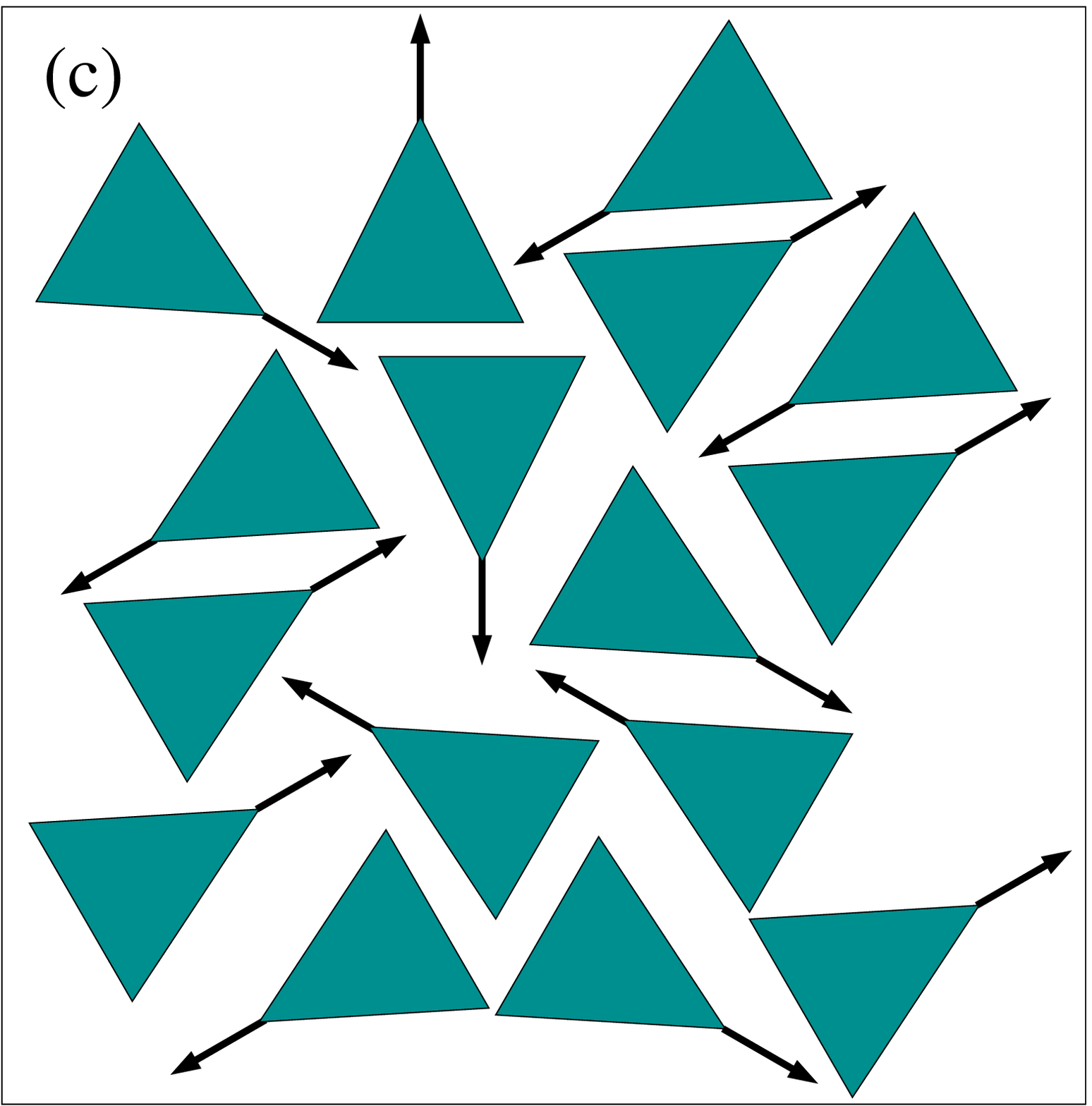,width=1.in}
	\caption{Sketch of I (a), N (b) and TR (c) uniform phases of HT (two latters show 
	perfect orientational ordering.}
	\label{lacero}
\end{figure}

This work is focused on the 
relative stability between the uniform phases of HT: I, N and TR phases sketched in Fig \ref{lacero}. 
In 2D the effect of three- and higher-order correlations becomes crucial to adequately 
predict the phase behavior of hard particles, especially
 at high densities. As shown in our previous work \cite{MR2}, the 
I-T transition densities in a HR fluid 
dramatically decrease when third-virial contributions are taken into account.
We expect the same effect here: inclusion of three-body correlations  
will certainly enhance the stability of the TR phase with respect to non-uniform 
phases (the latter are not included in the present study), as predicted by
simulations \cite{Dijkstra}. Also, higher-order correlations can even change the symmetry of the 
stable phases of HT for specific geometries. For example, right-angled triangles exhibit a stable 
rhombic phase \cite{Dijkstra}, while SPT predicts N phase stability instead. 
Despite these differences, we believe a second-virial-based theory is still valuable  
to systematically study the role of two-body interactions in promoting entropically-driven phase 
transitions in the HT fluid. These interactions predict a N-N first-order 
phase transition even for the one-component fluid. Also the I-N transition 
is of first order for some aspect ratios. 
Interestingly, simulations on right-angled 
isosceles triangles also predict a first-order I-R nematic phase transition 
\cite{Dijkstra}. 

The region of stability of the TR phase predicted by SPT extends away from an aspect ratio of
$\kappa=\sqrt{3}$, and also includes isosceles triangles with wider or more acute 
opening angles. The second-order I-N transition in the hard needle limit 
$\kappa\to\infty$ (for acute triangles) or $\kappa\to 0$ (for obtuse triangles) 
result in different asymptotic expressions for the
 packing fractions as a function of $\kappa$. While at both limits the packing fraction 
 $\eta$ goes to zero   
as $\kappa^{-1}$ (for acute HT) or $\kappa$ (for obtuse HT) the coefficients of proportionality are 
different. We also study the phase
behavior of symmetric binary mixtures, with species having the same areas and with 
aspect ratios symmetric with respect to the equilateral triangle: 
$\kappa_1\kappa_2=3$. We found that the mixing of certain symmetric triangles 
can stabilize the TR phase, which is not stable in the two one-component limits. The 
phase diagram (in the pressure-composition plane) exhibits
second- and first-order I-N transitions, a first-order 
N-N transition ending in the critical point and, at high pressures,
a Landau point where two first-order N-TR transitions and a N-N 
demixing transition coalesce, resembling the phase-diagram topology of 
symmetric rod-plate mixtures in 3D \cite{roij,yuri_prl}.

The article is organized as follows. Section \ref{spt} is devoted to present 
the SPT for general mixtures of HT. Analytical 
expressions for the I-N or I-TR second-order transition lines resulting from 
a bifurcation analysis are presented, along with some details on the numerical  
minimization of the free-energy with respect to the coefficients of 
the Fourier expansion of $f_i(\phi)$. In Sec. \ref{excluded} expressions for the 
excluded area of HT mixtures and for the Fourier coefficients, the key ingredients of the theoretical 
model, are presented. Sec. \ref{one-component} is devoted to the 
study of the one-component case. We begin the section by showing the symmetries of the 
excluded area, which allow us to explain some of the orientational ordering 
properties of HT. The complete 
phase diagram of HT (including only uniform phases) is analyzed in detail.
In Sec. \ref{binary} phase diagrams are presented for 
symmetric and asymmetric binary mixtures, the latter exhibiting  
a strong I-N demixing transition. Finally some conclusions are drawn in Sec. 
\ref{conclusions}. 

\section{SPT model for a binary mixture of convex bodies}
\label{spt}
This section is devoted to presenting expressions for the free-energy of a binary mixture 
of HT using the SPT formalism. The one-component case 
is trivially obtained by fixing the molar fraction of the corresponding species equal to unity. 
The binary mixture is described in terms of the one-body 
density distribution function of species $i$ ($i=1,2$), $\rho_i(\phi)$, which can be written as
\begin{eqnarray}
\rho_i(\phi)=\rho_i f_i(\phi),
\end{eqnarray}
with $\rho_i=\rho x_i$ the number density of species $i$, defined as the product of 
the total number density $\rho$ (with $\rho=\sum_i\rho_i$) 
and its molar fraction $x_i$ (which fulfills the constraint $\sum_i x_i=1$). 
The function $f_i(\phi)$ is 
the orientational distribution function of species $i$, which satisfies 
$\int_0^{2\pi}d\phi f_i(\phi)=1$. The total packing fraction of the mixture is calculated as
\begin{eqnarray}
\eta=\sum_i \rho_i a_i,
\end{eqnarray}
with $a_i$ the particle area of the corresponding species. According to SPT,
the excess part of the free-energy density of a binary mixture of convex 2D bodies
reads \cite{yuri_demix,MR2,dani_demix} 
\begin{eqnarray}
\Phi_{\rm ex}[\{\rho_i\}]\equiv \frac{\beta {\cal F}_{\rm ex}[\{\rho_i\}]}{A}
	=-\rho\log\left(1-\eta\right)+\frac{\rho^2{\cal K}\left[\{\rho_i\}\right]}{1-\eta}, 
\end{eqnarray}
with $\beta=\left(k_B T\right)^{-1}$ the inverse temperature, $A$ the total area 
of the system, 
and ${\cal F}_{\rm ex}[\{\rho_i\}]$ the excess part of the free-energy density functional. We have 
defined  
\begin{eqnarray}
	{\cal K}\left[\{\rho_i\}\right]&&\equiv \frac{1}{2}
\sum_{i,j}x_ix_j\int d\phi_1\int d\phi_2 f_i(\phi_1)f_j(\phi_2)\nonumber\\
		&&\times A^{(\rm spt)}_{ij}(\phi_{12}),
\end{eqnarray}
where $A^{(\rm spt)}_{i,j}(\phi)$ is related to the excluded area, 
$A^{(\rm excl)}_{ij}(\phi)$, 
between species $i$ and $j$, with a relative angle between their axes equal to $\phi$. 
The corresponding relation is  
\begin{eqnarray}
A^{(\rm spt)}_{ij}(\phi)=A^{(\rm excl)}_{ij}(\phi)-a_i-a_j.
\end{eqnarray}
As usual, the ideal part of the free-energy density is calculated as  
\begin{eqnarray}
	\Phi_{\rm id}[\{\rho_i\}]&&\equiv \frac{\beta {\cal F}[\{\rho_i\}]}{A}\nonumber\\
	&&=\sum_i \int d\phi \rho_i(\phi)
\left[\log\left (\rho_i(\phi){\cal V}_i\right)-1\right],
\end{eqnarray}
with ${\cal V}_i$ the thermal area of species $i$. The total free-energy density is simply  
the sum $\Phi[\{\rho_i\}]=\Phi_{\rm id}[\{\rho_i\}]+\Phi_{\rm ex}[\{\rho_i\}]$.
It is useful to express the orientational distribution function $f_i(\phi)$ in terms 
of its Fourier expansion:
\begin{eqnarray}
f_i(\phi)\equiv\frac{1}{2\pi}\Psi_i(\phi)=\frac{1}{2\pi}
\left[1+\sum_{k\geq 1} f_k^{(i)}\cos (k\phi)\right],
\end{eqnarray}
with $\{f_k^{(i)}\}$ the Fourier amplitudes. Using this expansion the total free-energy 
per particle  in reduced thermal units
\begin{eqnarray}
\varphi[\{\rho_i\}] \equiv \frac{\Phi[\{\rho_i\}]}{\rho},
\end{eqnarray}
can be written as
\begin{eqnarray}
&&\varphi=\log y-1+\sum_i x_i\left\{\log x_i
\right.\nonumber\\
	&&\left.+\frac{1}{2\pi}\int_0^{2\pi}d\phi \Psi_i(\phi)\log \Psi_i(\phi)\right\}+y 
	{\cal K}(\{f_k^{(i)}\}),
	\label{varphi}\\
	&&{\cal K}(\{f_k^{(i)}\})=\frac{1}{2}\sum_{i,j}x_ix_j\sum_{k\geq 0}s_k^{(i,j)}f_k^{(i)}f_k^{(j)},
\label{varphi_1}
\end{eqnarray}
where $f_0^{(i)}=1$ and the constant terms $2\pi {\cal V}_i$ have been dropped. Also we defined
$\displaystyle{y\equiv\frac{\rho}{1-\eta}}$ and the coefficients 
\begin{eqnarray}
s_k^{(i,j)}=\frac{1}{2\pi}\int_0^{2\pi} d\phi A^{(\rm spt)}_{ij}(\phi)\cos(k\phi).
\label{coefficients}
\end{eqnarray}
According to SPT \cite{yuri_demix}, the pressure is calculated from
\begin{eqnarray}
	\beta p=\frac{\partial \Phi_{\rm ex}}{\partial \eta}=y+y^2{\cal K}(\{f_k^{(i)}\}),
\end{eqnarray}
and the Gibbs free-energy per particle in reduced thermal units, 
for a fixed value of the pressure $p_0$, is 
\begin{eqnarray}
g=\varphi+\beta p_0\left(\langle a\rangle +y^{-1}\right),\quad 
\langle a\rangle=\sum_i x_i a_i. 
\label{la_g}
\end{eqnarray}
From the equality $p=p_0$ we obtain 
\begin{eqnarray}
	y=\frac{\sqrt{1+4\beta p_0{\cal K}(\{f_k^{(i)}\})}-1}{
2T(\{f_k^{(i)}\})}. \label{la_y}
\end{eqnarray}
The partial derivatives of the Gibbs free-energy per particle, $g$, with respect to 
the Fourier coefficients $f_k^{(i)}$, are given by
\begin{eqnarray}
	\frac{\partial g}{\partial f_k^{(i)}}&&=
x_i\left\{\frac{1}{2\pi}\int_0^{2\pi}d\phi \cos(k\phi)\log \Psi_i(\phi)\right.\nonumber\\
	&&\left.+y\sum_j x_j s_k^{(i,j)}f_k^{(j)}\right\},\quad k\geq 1.
\label{gradient}
\end{eqnarray}
Defining $x\equiv x_1$ (so that $x_2=1-x$), we fix $p_0$ and, from 
(\ref{la_y}), obtain $y(x)$. Substitution in (\ref{la_g}) 
provides the Gibbs free energy, $g(x)$, as function of the mixture composition. 
The values of the Fourier coefficients $\{f_k^{(i)}\}$ are those which 
minimize the Gibbs free-energy $g$. A 
conjugate-gradient minimization was implemented, with analytical gradients calculated from 
(\ref{gradient}) and using a number of Fourier components ranging from 20 to 50 for 
each species. Their precise values depend on how sharp the functions $f_i(\phi)$ are. 
The common-tangent construction on $g(x)$ gives the coexistence conditions.

Alternatively, coexistence can be calculated through the equality of the
chemical potential of species $i$ belonging to different coexisting phases. These are 
calculated from
\begin{eqnarray}
	\beta \mu_i&&= \log\left(yx_i\right)+\frac{1}{2\pi}\int_0^{2\pi}
d\phi \Psi_i(\phi)\log \Psi_i(\phi)\nonumber\\
	&&+y\sum_j x_j \sum_k s_k^{(i,j)}
f_k^{(i)}f_k^{(j)}+\beta p_0a_i, 
\end{eqnarray}
where $p_0$ is a fixed value of pressure, and $y$ is calculated from
(\ref{la_y}), while $\{f_k^{(i)}\}$ are calculated from Gibbs free-energy minimization.
The degree of orientational order of species $i$ is measured by the N ($k=2$) and 
TR ($k=6$) order parameters: 
\begin{eqnarray}
Q_k^{(i)}=\int_0^{2\pi} d\phi f_i(\phi)\cos(k\phi),\quad k=2,\ 6.
\end{eqnarray} 

The packing fraction of the
second-order I-(N,TR) transition can be calculated by expanding the expression
$\displaystyle{\frac{\partial g}{\partial f_k^{(i)}}}$
given by Eq. (\ref{gradient}) up to first
order in $f_k^{(i)}$
($k=2$ for the N and $k=6$ for the TR phase). The equilibrium condition is obtained by
equating the result to zero, which can be written in matrix form as
\begin{eqnarray}
R{\bm f}_k={\bm 0}.
\label{lineal}
\end{eqnarray}
Here we defined the vector
${\bm f}_k\equiv \left(f_k^{(1)},f_k^{(2)}\right)^{\rm T}$
while the $R$-matrix elements are
\begin{eqnarray}
R_{ij}\equiv \delta_{ij}+2yx_js_k^{(i,j)},
\end{eqnarray}
with $\delta_{ij}$ the Kronecker delta.
The linear system (\ref{lineal}) has a non-trivial solution if
$\text{det}\left(R\right)=0$, which implies that the packing fraction at the transition is
\begin{eqnarray}
\eta^*=\frac{
2+\Delta_0-\sqrt{\Delta_0^2-4\Delta_1}}
{2\left(1+\Delta_0+\Delta_1\right)},
	\label{spinodal}
\end{eqnarray}
where we have defined
\begin{eqnarray}
&&\Delta_0=-\frac{2}{\langle a\rangle}
	\left(x_1s_k^{(1,1)}+x_2s_k^{(2,2)}\right), \label{delta_0}\\
&&\Delta_1=\frac{4x_1x_2}{\langle a\rangle^2}
\left[s_k^{(1,1)}s_k^{(2,2)}-\left(s_k^{(1,2)}\right)^2\right].
\label{delta_1}
\end{eqnarray}

\section{Excluded area between two HT}
\label{excluded}

\begin{figure}
\epsfig{file=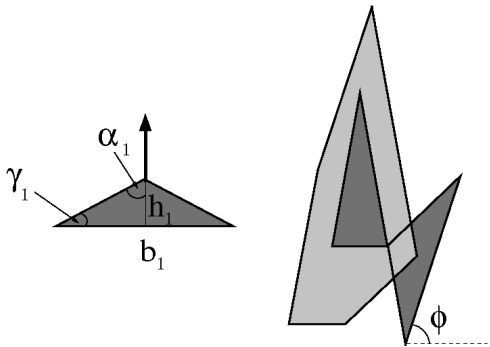,width=2.5in}
\caption{Sketch of the excluded area between two HT with relative 
angle between their axes equal to $\phi$. The main  
parameters (base, height, angles) used to define the geometry 
of the triangle are labeled.}
\label{fig1}
\end{figure}

The excluded area between two isosceles triangles of base $b_i$  
and heights $h_i$ ($i=1,2$) with main axes at a relative angle $\phi$ is sketched in 
Fig. \ref{fig1}. The main axis is defined in the direction of the particle height, i.e. 
pointing from the base to the opposite vertex. The aspect ratio of species $i$ 
is $\kappa_i=2h_i/b_i$, and $\alpha_i=\arctan\left(\kappa^{-1}_i
\right)$ is the angle between the height and the equally sized edge-lengths.
$\gamma_i=\pi/2-\alpha_i=\arctan(\kappa_i)$ is the angle between the latter and the base.
We also define the angles
$\alpha_{ij}=\alpha_i+\alpha_j$
and $\displaystyle{\theta_{ij}=\arctan\left[\frac{1}{2}\left(\kappa_i+\kappa_j
\right)\right]}$. For the case where the aspect ratios
fulfill the relation $\kappa_i\kappa_j\geq 3$, 
the functions $A_{ij}^{(\rm spt)}(\phi)$ can be calculated as
\begin{eqnarray}
	&&A_{ij}^{(\rm spt)}(\phi)\nonumber\\
	&&	=\left\{ 
\begin{matrix}
	{\cal B}_{ij}\cos\phi, & 
0\leq \phi\leq \alpha_{ij},\\
	\frac{{\cal B}_{ij}}{2}\cos\phi
	+\left(h_ih_j-\frac{b_ib_j}{4}\right)\sin\phi, & 
\alpha_{ij}\leq \phi\leq \theta_{ij},\\
	-\frac{{\cal B}_{ij}}{2}\cos\phi
	+\left(h_ih_j-\frac{3b_ib_j}{4}\right)\sin\phi, & 
\theta_{ij}\leq \phi\leq \pi,
\end{matrix}
\right.
\nonumber\\
	&&
\label{excluded1}
\end{eqnarray}
where we have defined ${\cal B}_{ij}=b_ih_j+b_jh_i$.
For $\kappa_i\kappa_j\leq 3$ we obtain: 
\begin{eqnarray}
	&&A_{ij}^{(\rm spt)}(\phi)\nonumber\\
	&&=\left\{ 
\begin{matrix}
	{\cal B}_{ij}\cos\phi, & 
0\leq \phi\leq \theta_{ij},\\
b_ib_j\sin\phi, &
\theta_{ij}\leq \phi\leq \alpha_{ij},\\
	-\frac{{\cal B}_{ij}}{2}\cos\phi
	+\left(h_ih_j-\frac{3b_ib_j}{4}\right)\sin\phi, & 
\alpha_{ij}\leq \phi\leq \pi,
\end{matrix}
\right.\nonumber\\
	&&
\label{excluded2}
\end{eqnarray}
For angles $\pi<\phi\leq 2\pi$ this function can be calculated by invoking its
symmetry with respect to $\pi$:  
$A_{ij}^{(\rm spt)}(\pi+\phi)=
A_{ij}^{(\rm spt)}(\pi-\phi)$ (with $\phi\in[0,\pi]$).

Defining the angles $\gamma_{ij}^{\pm}=\gamma_i\pm \gamma_j$, the coefficients 
$s_k^{(i,j)}$ in (\ref{coefficients}) can be calculated as
\begin{eqnarray}
&&s_1^{(i,j)}=\frac{c_{ij}}{2}\left(\theta_{ij}-\frac{\gamma_{ij}^+}{2}\right)\sin\gamma_{ij}^+,\\
	&&s_k^{(i,j)}\nonumber\\
	&&=-\frac{c_{ij}2^{\delta_{k0}-1}}{k^2-1}\left\{
(-1)^k\left[\cos\left(k\gamma_{ij}^+\right)+2\cos\gamma_{ij}^-
	+\cos\gamma_{ij}^+\right]\right.\nonumber\\
	&&\left.+2\sqrt{1+2\cos\gamma_{ij}^-\cos\gamma_{ij}^++\cos^2\gamma_{ij}^-}\cos(k\theta_{ij})\right\},
\nonumber\\
	&&
\end{eqnarray}
for $k\neq 1$, where 
\begin{eqnarray}
&& c_{ij}=\frac{l_i l_j}{\pi}=
\frac{1}{\pi}\sqrt{a_ia_j}
\times \sqrt{\left(\kappa_i+\frac{1}{\kappa_i}\right)
\left(\kappa_j+\frac{1}{\kappa_j}\right)},\nonumber\\
	&&
\end{eqnarray}
with $l_i=\sqrt{h_i^2+(b_i/2)^2}$ the equally-sized lengths of triangle $i$,
and $a_i=b_i h_i/2$ its particle area.

\section{One-component fluid}
\label{one-component}

\begin{figure}
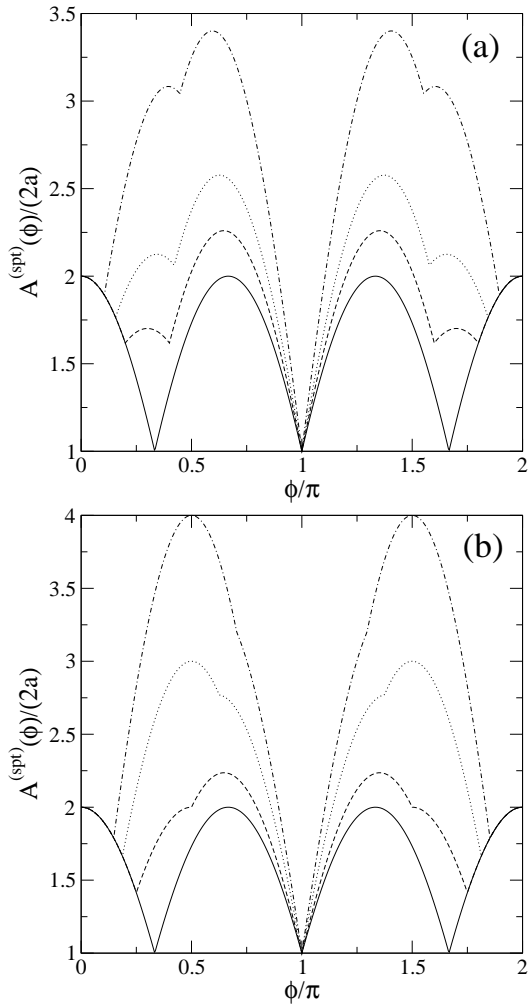

\epsfig{file=fig3a.eps,width=2.7in}
\epsfig{file=fig3b.eps,width=2.7in}
\caption{(a) The function $A^{(\rm spt)}(\phi)/(2a)$ for triangles 
with $\kappa\geq \sqrt{3}$.
Chosen values are $\kappa=\sqrt{3}$ (solid), $\sqrt{5+2\sqrt{5}}$ (dashed),
4 (dotted) and 6 (dot-dashed). (b) The same function for triangles with $\kappa\leq \sqrt{3}$ and
values: $\sqrt{3}$ (solid), 1 (dashed), 2/3 (dotted) and 1/2 (dot-dashed).}
\label{fig2}
\end{figure}

Fig. \ref{fig2} shows the function $A^{(\rm spt)}(\phi)\equiv A_{11}^{(\rm spt)}(\phi)$,
scaled by $2a$ ($a=a_1$), for four different
values of aspect ratio corresponding to triangles with $\kappa\geq \sqrt{3}$ (a), and 
$\kappa_1\leq \sqrt{3}$ (b). For the sake of brevity, in this section we drop the species 
labels in all the magnitudes defined in the preceding section.

As can be seen from the figure, this function has a rather complex form: it may have up to three 
local minima and three local maxima in the interval $[0,\pi]$. The absolute minimum is reached 
at $\phi=\pi$, for which the excluded volume is $A^{(\rm excl)}=4 a$, while for $\phi=0$ 
the function $A^{(\rm spt)}(\phi)$ has a local maximum corresponding to 
$A^{(\rm excl)}=6 a$. The absolute maximum is always located in the interval $[\pi/2,\pi)$ 
and its position tends to $\pi/2$ as the aspect ratio moves away from $\sqrt{3}$ (the 
value corresponding to the equilateral triangle) in both 
directions: $\kappa\to \infty$ or $\kappa\to 0$. It is interesting to note that for 
$\kappa=\sqrt{5+2\sqrt{5}}$, the aspect ratio corresponding to the sublime triangle (that 
for which the ratio between the equally sized-lengths and the base of the triangle 
is just the golden ratio), two of the local minima of $A^{(\rm spt)}(\phi)$ 
have exactly the same value. As we will see later, the presence of local minima in the 
excluded area imposes some important symmetries in the orientational distribution function $f(\phi)$.
For equilateral triangles ($\kappa=\sqrt{3}$) the excluded area has a sixfold symmetry: the 
local maxima and local minima have the same heights and the former are located at $n\pi/3$ 
with $n=0,1,\dots,6$. This symmetry forces
the orientational distribution function of the nematic phase to have a periodicity of $\pi/3$: 
$f(\phi)=f(\phi+\pi/3)$. This phase, called the TR phase, has six equivalent nematic directors.

For the one component case (dropping the species indexes), the expressions for the coefficients 
(\ref{coefficients}) are more conveniently written as 
\begin{eqnarray}
	&&s_0=
	\frac{8l^2}{\pi}\cos^4\left(\frac{\gamma}{2}\right), \\
	&&s_{2j}\nonumber\\&&=-\frac{4l^2}{\pi(4j^2-1)}\left\{
\cos\left[\frac{(2j-1)\gamma}{2}\right]
	\cos\left[\frac{(2j+1)\gamma}{2}\right]\right\}^2, \nonumber\\&&\\
&&s_{2j+1}=
\frac{l^2}{\pi j(j+1)}\left\{\sin(j\gamma)\sin[(j+1)\gamma]
\right\}^2, 
\end{eqnarray}
with $j\geq 1$, while $s_1=0$.
The fact that the the even coefficients $s_{2j}$ are always negative,
while the odd ones are positive, has an important 
implication on the orientational properties of HT, namely, the absence of a polar N 
phase, as we prove in Appendix \ref{app1}. 



For the one-component fluid the packing fraction corresponding to the second order I-(N,TR) 
transition can be obtained from (\ref{spinodal})-(\ref{delta_1}) by taking $x_1=1$. The result is 
\begin{eqnarray}
	\eta_n^*&&=\left(1+\Delta_0\right)^{-1}=
\left[1+\frac{2}{(n^2-1)\pi}
\left(\kappa+\kappa^{-1}\right)\right.\nonumber\\
	&&\left.\times\left(\cos(n\gamma)+\cos\gamma\right)^2
\right]^{-1}, \quad n=2,6.
\label{etak}
\end{eqnarray}
For the particular case $n=2$ (corresponding to the I-N bifurcation), 
using $\tan\gamma=\kappa$, this can be rewritten as
\begin{eqnarray}
	\eta_2^*&&=\left[1+\frac{2}{3\pi}
\left(\kappa+\kappa^{-1}\right)\left(1-\frac{2}{\sqrt{\kappa^2+1}}\right)^2\right.\nonumber\\
	&&\left.\times\left(1+\frac{1}{\sqrt{\kappa^2+1}}\right)^2\right]^{-1}.
\end{eqnarray}
It is easy to see that this expression is not symmetric with respect to 
the change $\kappa\to\kappa^{-1}$. The limits $\kappa\to \infty$ and $\kappa\to 0$ 
(both corresponding to the same hard-needle limit) are not equivalent. Asymptotically we obtain 
$\displaystyle{\eta_2^*\sim \frac{3\pi}{2\kappa}}$ for the former and
$\displaystyle{\eta_2^*\sim  \frac{3\pi\kappa}{8}}$ for the latter. 
The TR phase is expected to be stable in some interval of aspect ratios $[\kappa_1^*,\kappa_2^*]$,
including the value for the equilateral triangle ($\kappa=\sqrt{3}$), for which TR is the only possible 
orientationally-ordered uniform phase. An estimation of this interval is obtained 
by solving the equality $\eta_2^*(\kappa)=\eta_6^*(\kappa)$ for $\kappa$. At these
points, the I-N and I-TR second-order transition curves intersect, the 
latter being below the former for aspect ratios values in the interval 
$[\kappa_1^*,\kappa_2^*]$. From (\ref{etak}), this equation is equivalent 
to finding the values of $\gamma$ ($\kappa=\tan\gamma$) for which 
\begin{eqnarray}
r\cos(2\gamma)+(r-1)\cos(\gamma)=\cos(6\gamma),
\end{eqnarray}
with $\displaystyle{r=\pm \sqrt{\frac{35}{3}}}$. This gives the 
values $\kappa_1^*=1.278$ and $\kappa_2^*=2.405$. 
However, this result is correct only if the 
involved phase transitions are of second order, which is not the case as we 
will promptly see. 

\begin{figure}
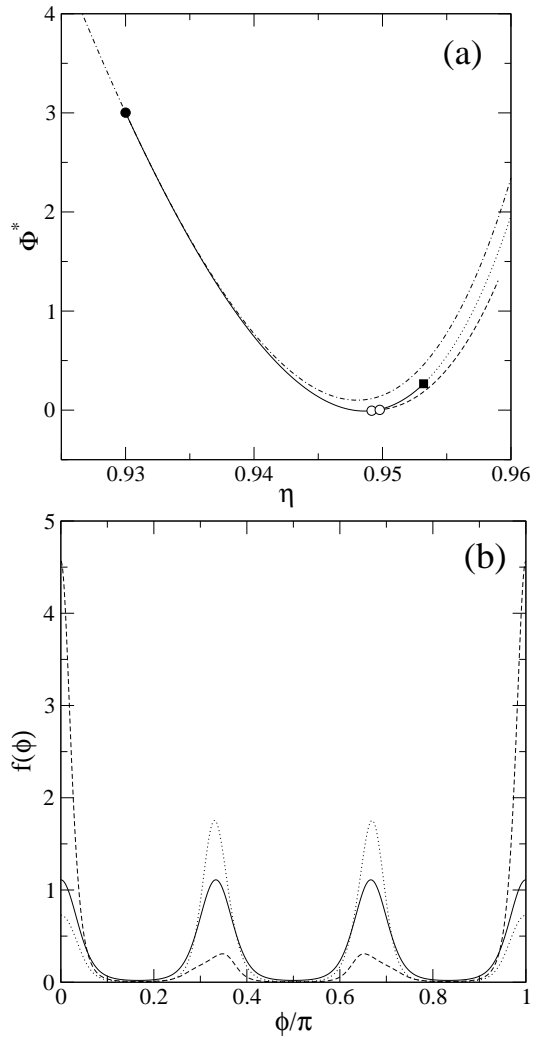

\epsfig{file=fig4a.eps,width=2.7in}
\epsfig{file=fig4b.eps,width=2.7in}
\caption{(a) Free-energy densities minus a straight line: 
$\Phi^*\equiv \Phi-a\eta-b$ (with $a=623.45$ and $b=-562.94$) corresponding to the I 
(dot-dashed), TR (solid), N (dashed) and TR$^*$ (dotted) phases of HT with $\kappa=2.2$. 
The solid circle indicates the I-TR bifurcation, while open circles represent the TR-N coexistence 
(with $\eta_{\rm TR}=0.949157$ and $\eta_{\rm N}=0.949785$) and the square shows 
the TR-TR$^*$ bifurcation. 
(b) Orientational distribution functions corresponding to TR (solid) at $\eta=0.949$, 
N (dashed) at $\eta=0.95$, both for stable phases of HT with $\kappa=2.2$. 
Dotted line is a distribution function corresponding to a TR$^*$ metastable phase with
$\eta=0.957$.}  
\label{fig3_new}
\end{figure}

As already pointed out, the TR phase is also stable for aspect ratios in the neighborhood 
of $\sqrt{3}$, corresponding to the equilateral triangle aspect ratio. 
The sequence of stable phases for $\kappa=2.2>\sqrt{3}\approx 1.73$ 
as $\eta$ is increased is shown in Fig. \ref{fig3_new} (a), where the 
free-energy densities of different stable and metastable branches are plotted. 
The I phase is stable up to 
$\eta\approx0.93$, at which the TR phase bifurcates through a second-order I-TR transition. 
The latter is stable up to $\eta_{\rm TR}=0.949157$, 
the coexisting value of the first-order TR-N transition. The orientational distribution 
function of the TR phase for $\eta=0.949$ (just below the TR-N transition) is shown 
in Fig. \ref{fig3_new} (b); it has the expected symmetry $f(\phi+\pi/3)=f(\phi)$. 
Also we show the orientational distribution function of the N phase for $\eta=0.95$ 
(above the transition). We notice the strong uniaxial ordering of HT at $\phi=\{0,\pi\}$,
although the presence of small peaks at $\phi\approx \{\pi/3,2\pi/3\}$ can be also seen.
From $\eta_{\rm N}=0.949785$ onwards the N phase is the stable phase up to close packing. 
It is interesting to see that the metastable TR phase changes continuously at $\eta=0.9532$ 
to a different asymmetric TR$^*$ phase [see panel (b)], with two sharp, equivalent 
peaks at $\phi\approx \{\pi/3,2\pi/3\}$, while the other two peaks at $\phi=\{0,\pi\}$ have 
considerably lower heights. This TR$^*$ phase is always metastable with respect to the 
usual uniaxial N phase. The same sequence of stable and metastable phases is found 
for $\kappa\lesssim\sqrt{3}$. We note here that MC simulations of equilateral HT found the I-TR 
transition at $\eta\approx 0.7$ with TR phase being stable up to $\eta\approx 0.89$ where 
a transition to a chiral triangular crystal takes place \cite{Dijkstra}. In this sense 
the I-(TR,N) transition packing fractions estimated by SPT for $\kappa\sim \sqrt{3}$ are clearly 
overestimated. We expect that the inclusion of three-body correlations in a modified theory will 
correct the transition densities as we have already done for the HR fluid \cite{MR2}.

\begin{figure}
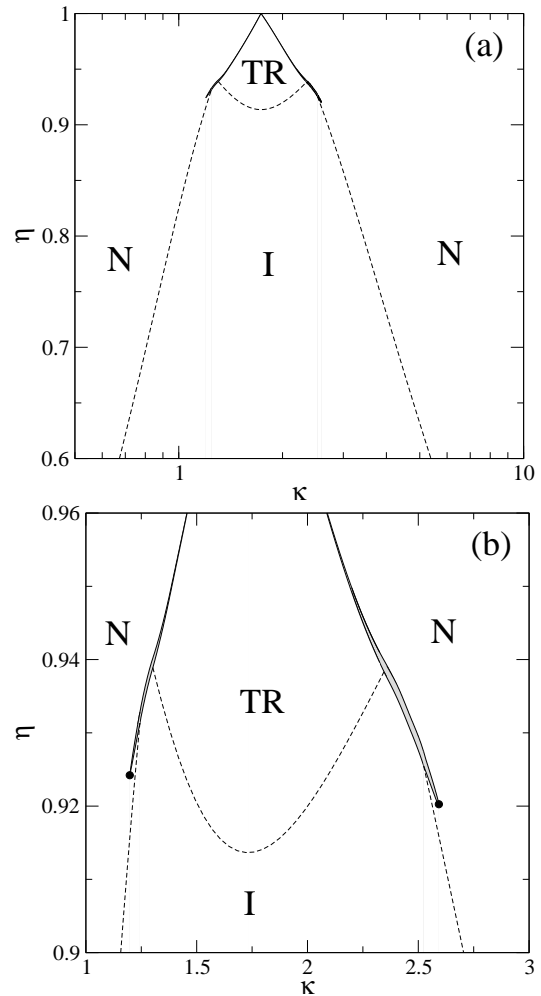

\epsfig{file=fig5a.eps,width=2.7in}
\epsfig{file=fig5b.eps,width=2.7in}
\caption{Phase diagram of HT in the packing-fraction ($\eta$)-aspect 
ratio ($\kappa$) plane. 
(a) $\eta\in[0.6,1]$ in logarithmic scale for $\kappa$. (b) Zoom of 
the phase diagram (with $\kappa$ in linear scale), showing 
the presence of N-N transitions ending in critical points (solid circles), the first-order character of the 
I-N transition for certain values of $\kappa$, and the region of stability of the TR phase. Dashed and solid 
lines show second- and first-order transitions, respectively. Regions 
of stability of different phases are correspondingly labeled.}
\label{fig3}
\end{figure}

The complete phase diagram of HT is shown in Fig. \ref{fig3}. The second-order transitions are 
shown with dashed lines, while the binodals of the first-order transitions are shown with 
solid lines. The first order transitions are so weak that the coexistence regions are not visible 
in the scale of the graph. As pointed out before, the asymmetric character of the second-order I-N transitions, as 
$\kappa\to \infty$ and $\kappa\to 0$, is apparent. The TR phase is stable in an interval 
of aspect ratios about $\sqrt{3}$, but the interval is now modified with respect to the values obtained 
from the bifurcation analysis due to the first-order character of the I-N and TR-N transitions for 
some values of $\kappa$. Now the extrema of this interval are $\kappa_1^*=1.30$ and $\kappa_2^*=2.35$, 
the values corresponding to the critical end-points where the second-order I-TR transition line meets  
the I (from below) and TR (from above) binodals of the I-N and TR-N first-order transitions, 
respectively. The TR phase is bounded from above by the TR-N first-order transition, although we cannot 
discard the presence of tricritical points for values of $\kappa$ close to $\sqrt{3}$, where 
the TR-N transition would become of second order. 
The numerical procedure used to calculate the TR-N coexistence becomes unstable for values of 
$\eta$ close to 1, so we have extrapolated the binodals up to close packing, $\eta=1$.

But the most striking feature of the phase diagram is the presence of first-order N-N transitions 
at both sides of $\sqrt{3}$, both ending in critical points $\kappa_1^c\approx 1.2$ and 
$\kappa_2^c\approx 2.6$ [see panel (b)]. 
When changing these values of aspect ratios to $\kappa=\sqrt{3}$, these N-N transitions meet the 
second-order I-N transition at two critical end-points, and from these points we find the 
first-order I-N transitions already described. To the best of our knowledge this is the first example
where a N-N transition is found for a one-component hard-particle system.

\begin{figure}
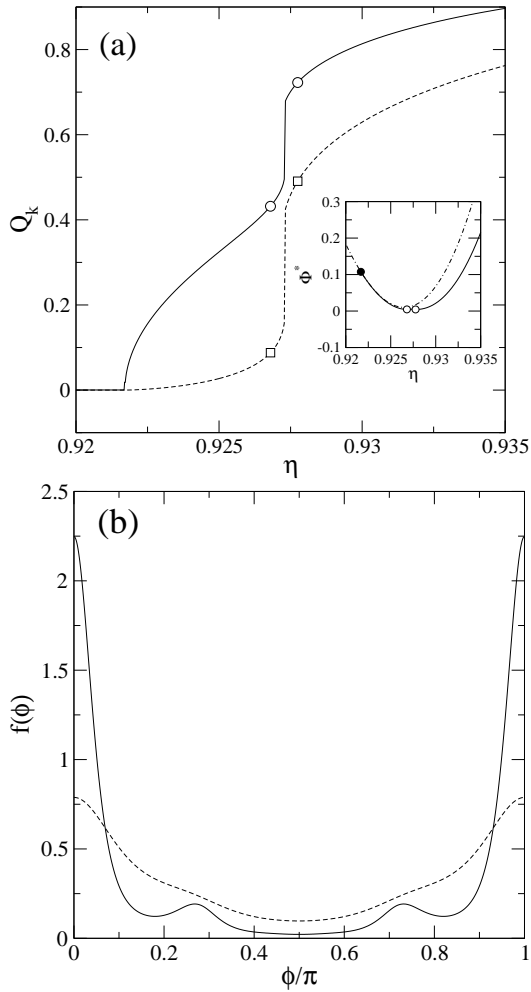

\epsfig{file=fig6a.eps,width=2.7in}
\epsfig{file=fig6b.eps,width=2.7in}
\caption{(a) Order parameters $Q_k$ (solid for $k=2$ and dashed for $k=6$) as a function 
of the packing fraction $\eta$ for HT with $\kappa=1.215$, showing the N$_1$-N$_2$ first-order 
transition (shown with circles and squares, respectively). Inset: free-energy densities 
minus a straight line, $\Phi^*=\Phi-a\eta-b$ (with $a=-288.33$ and $b=334.6$), corresponding to the I 
(dot-dashed) and N$_i$ (solid) phases as a function of $\eta$. The solid circles indicates the
I-N bifurcation point, while the open circles show the N-N coexistence values. (b) Orientational 
distribution functions corresponding to the coexisting N$_1$ (dashed) and N$_2$ (solid) phases 
for the packing-fraction values $\eta_1=0.92680$ and $\eta_2=0.92776$ respectively.}  
\label{fig5}
\end{figure}

Fig. \ref{fig5} (a) shows the order parameters $Q_k$ ($k=2,6$) as a function of $\eta$ 
for $\kappa=1.215$, a value for which a N-N transition takes place. We can see the abrupt changes
in order parameters 
as $\eta$ is increased. The coexistence values of $\eta$ are indicated on the curves as symbols,
while the inset shows the free-energy densities of the I and N phases, together with the coexistence
points found from the usual double-tangent construction. The orientational distribution functions 
corresponding to the coexisting N phases are plotted in panel (b). We notice the difference 
in orientational ordering: while the less dense N is sightly ordered, particles
in the other phase are strongly oriented along the uniaxial director. The presence of  
small peaks at $\phi\approx \{\pi/3,2\pi/3\}$ is due to the rather small orientational TR
correlations, as the aspect ratio is not too far from $\sqrt{3}$.    

\section{Binary mixtures}
\label{binary}

\begin{figure}
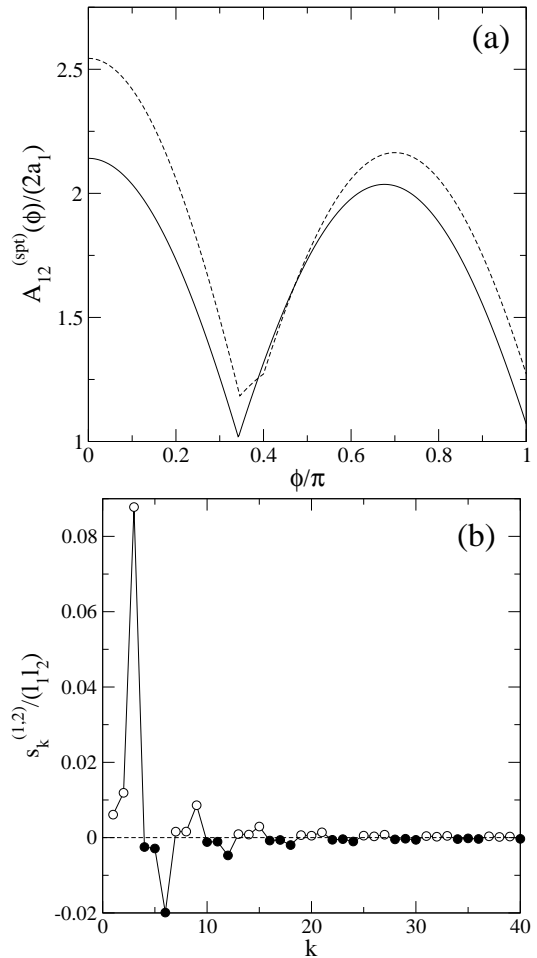

\epsfig{file=fig7a.eps,width=2.7in}
\epsfig{file=fig7b.eps,width=2.7in}
\caption{(a) The function $A_{12}^{(\rm spt)}(\phi)$ related to the 
excluded areas between two symmetric ($\kappa_1\kappa_2=3$) 
triangles of equal areas ($a_1=a_2$) and $r=b_2/b_1=1.453$ (solid), and   
between the sublime and Gnomon (with a ratio between its base and the equally-sized lengths 
equal to the golden number) triangles of equal areas (dashed).
(b) Scaled Fourier coefficients of the pair of species corresponding to the 
solid line in (a). Open and solid circles correspond to positive 
	and negative coefficients, respectively.}
\label{fig6}
\end{figure}

Now we present results obtained from the numerical implementation of 
SPT as applied to a particular symmetric mixture of HT. In this mixture, 
the aspect ratios of the two species
fulfill the condition $\kappa_1\kappa_2=3$ and have equal areas, $a_1=a_2$. 
The ratio between their bases is chosen to be $r=b_2/b_1=1.453$. Obviously the results will 
not depend on the particular value of one of the bases, so we set $b_1=1$. We
obtain $\kappa_1=\sqrt{3}r=2.52$ and $\kappa_2=\sqrt{3}/r=1.19$. Note that these values 
are outside the interval $[1.3,2.35]$ inside which the TR phase is stable in the one-component 
system. The chosen particle symmetry gives $\alpha_{ij}=\theta_{ij}$, 
so the interval $[\alpha_{ij},\theta_{ij}]$ in the expressions for $A^{(\rm spt)}_{ij}(\phi)$, 
given by Eq. (\ref{excluded1}) and (\ref{excluded2}), shrinks to zero and 
both expressions become equivalent. The consequence of this symmetry can be seen 
in Fig. \ref{fig6} (a), where we plot the functions $A^{(\rm spt)}_{12}(\phi)$ for the 
previously defined pair of symmetric triangles and also for a non-symmetric pair. 
Note how the vanishing of the interval $[\alpha_{ij},\theta_{ij}]$ makes the excluded area 
much more similar to that of equilateral triangles. As we will promptly see, this property has 
a profound impact on the relative stability of the TR phase in binary mixtures. The previously 
reported property of the one-component fluid, namely the positiveness (negativeness) of odd (even) 
Fourier coefficients, does not have a counterpart in binary 
mixtures, as can be seen from Fig. \ref{fig6} (b). Therefore, the previous proof on the absence 
of a polar phase in the one-component fluid of HT looses its validity for binary mixtures.
Because of this point, we were forced to use all the Fourier coefficients in the free-energy minimization. 
However, for all the pair of triangles explored, we always 
obtained free-energy minima with vanishing odd Fourier coefficients, so that the polar phase 
can probably be discarded.

\begin{figure}
\epsfig{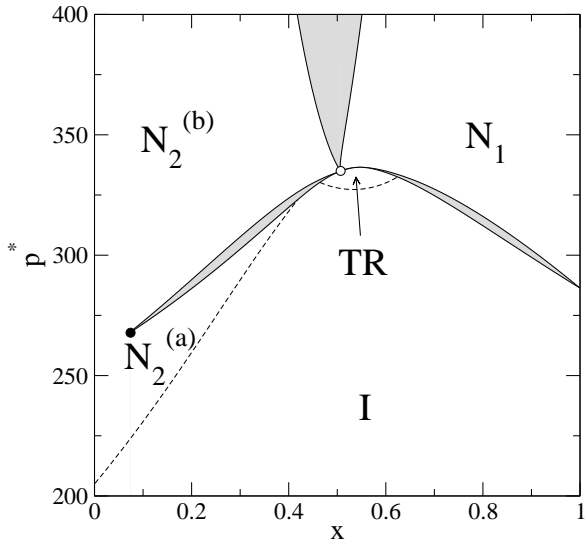}
\caption{Phase diagram in the scaled pressure ($p^*=\beta p a_1$)-molar 
fraction ($x\equiv x_1$) plane of the HT with $\kappa_1=2.52$ (species 1) 
and its symmetric ($\kappa_1\kappa_2=3$) counterpart, both of equal areas 
and with $r=b_2/b_1=1.453$. Regions of stability of different phases are 
correspondingly labeled. Dashed and solid lines represent the second-order 
transition curves and the binodals of first-order transitions, respectively. 
Coexistence regions are shaded in grey. The solid and open circles 
represent the critical and Landau points, respectively.}
\label{fig7}
\end{figure}

The phase diagram of the symmetric binary mixture ($\kappa_1\kappa_2=3$) with equal particle areas 
and $r=b_2/b_1=1.453$ is shown in Fig. \ref{fig7} in the reduced pressure-composition plane. 
Species 1, having aspect ratio $\kappa_1=2.52>\sqrt{3}$, corresponds to an acute-angled triangle 
with $\gamma_1>\pi/3$, while the other one, with $\kappa_2=1.19<\sqrt{3}$, is also an acute-angled 
triangle but with $\gamma_2<\pi/3$. A second-order I-N$_2$ transition line departs from the 
$x\equiv x_1=0$ axis, terminating at the critical end-point where it meets from below (above) the 
N$_2^{(a)}$ (I) binodal of the N$_2^{(a)}$-N$_2^{(b)}$ (I-N$_2^{(b)}$) transition.  
As pointed out before, the one-component fluid of HT with $\kappa<\sqrt{3}$ 
exhibits a N-N transition for $\kappa\in [1.2,1.24]$. The aspect ratio $\kappa_2=1.19$ of species 2
is below this interval, so this transition cannot end with $x=0$ but instead it ends in
a critical point. Also the one-component HT fluid with $\kappa>\sqrt{3}$ exhibits a N-N transition for 
$\kappa\in[2.52,2.60]$, and a I-N first-order transition for $\kappa\in[2.34,2.52]$, which turns into 
a TR-N transition for $\kappa<2.34$ [see Fig. \ref{fig3} (b)]. The aspect ratio $\kappa_1=2.52$ coincides with one of the 
limits of both intervals, so the first order I-N$_1$ transition in the binary mixture this time
ends at $x=1$. We found that the present mixture exhibits a second-order 
I-TR transition ending at the I binodal of the I$_2^{(a)}$-N$_2^{(b)}$ transition 
(on the left) and at the I binodal of the I-N$_1$ transition (on the right). Above this line there 
exists a region of TR phase stability, bounded above by the TR binodal of the first-order TR-N$_i$ 
transitions. Both transitions coalesce at 
a Landau point above which there appears a N$_1$-N$_2^{(b)}$ demixing transition with a demixed 
gap that increases with pressure. 
It is interesting to note the similarity between this phase diagram and that of a  
symmetric binary mixture of rods and plates in three dimensions (with species of equal volumes 
and aspect ratios satisfying $\kappa_1\kappa_2=1$): the first order I-N transitions departing 
from $x=0$ and $x=1$ end at the Landau point, above which there is either N-N demixing or a small 
region of biaxial nematic phase stability 
bounded above by N-N demixing. The most important conclusion that can be drawn from the topology of the 
phase diagram shown in Fig. \ref{fig7} is that by appropriately choosing the triangular geometries of 
the species (i.e. their symmetrization), the TR phase can be stabilized in mixtures of species 
which by themselves cannot stabilize this phase in their corresponding one-component phase diagrams.   

\begin{figure}
\epsfig{file=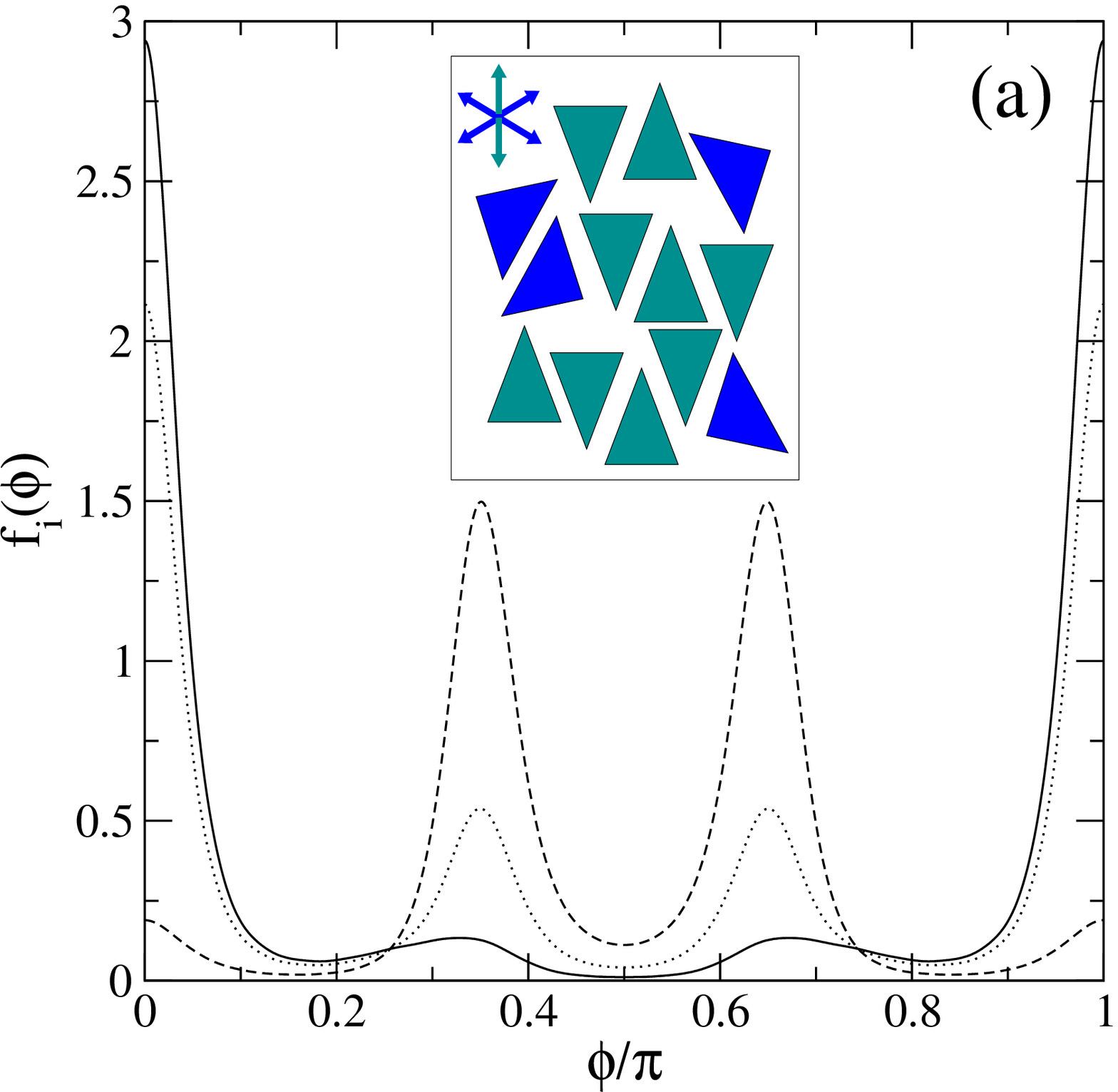,width=2.7in}
\epsfig{file=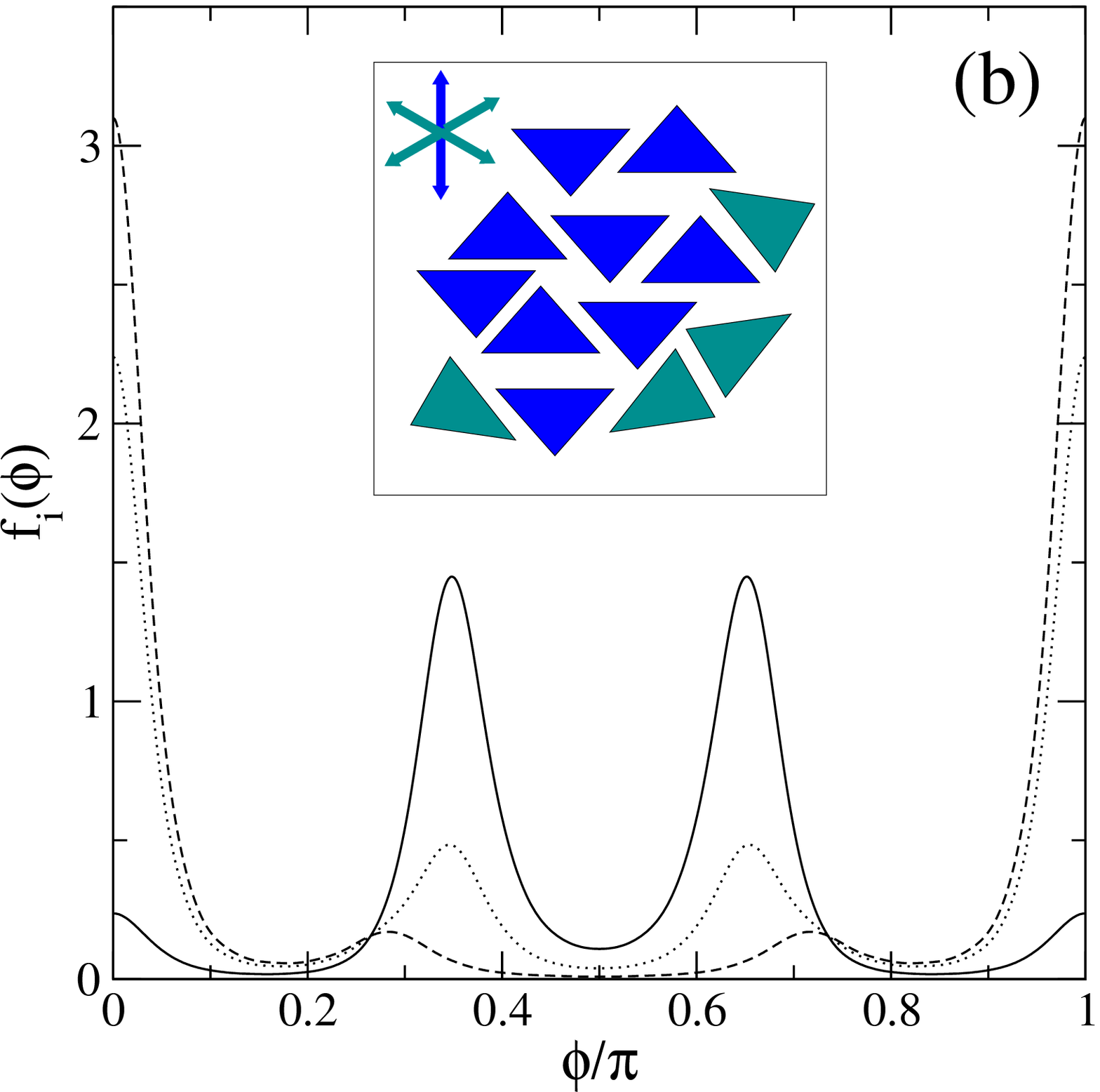,width=2.7in}
\caption{Distribution functions of the first (solid) and second (dashed) 
species of the same binary mixture as in Fig. \ref{fig7} for a fixed 
scaled pressure $p^*=346.042$ and molar fractions $x=0.7$ (a) and $x=0.3$ (b). Dotted lines indicate the
total orientational distribution function $f_{\rm t}(\phi)$ (see text for definition). Insets are 
	schematic configurations for binary mixtures corresponding to each of 
	subfigures (for simplicity perfect orientational ordering approximation was assumed 
	with arrows indicating the orientations of particle axes.}
\label{fig8}
\end{figure}

The orientational ordering of different species at pressures above the Landau point in the 
regions of N phase stability is shown in Fig. \ref{fig8}. 
At $x=0.7$ species 1, with a higher composition, possesses an orientational distribution function with a clear 
uniaxial nematic symmetry: two sharp peaks located at $0$ and $\pi$, with rather small TR correlations at $\pi/3$ 
and $2\pi/3$. But the axes of the second species exhibit two clear sharp peaks at $\sim\pi/3$ and $\sim 2\pi/3$, while 
their proportion for orientations along $0$ and $\pi$ is very small. This orientational segregation 
of species implies that the total orientational distribution function, defined as 
$f_{\rm t}(\phi)=\sum_i x_i f_i(\phi)$, has two main peaks located at $0$ and $\pi$, indicating the 
uniaxial character of the N phase; but strong TR correlations are present, as indicated by the fact that the 
other two orientations have well-developed peaks. 
A similar situation occurs at the other side of the phase diagram, for $x=0.3$. Now the second species has
a clear uniaxial N symmetry, while the axes of the first species point to the other two directions, $\{\pi/3,2\pi/3\}$, 
to give a total orientational distribution function $f_{\rm t}(\phi)$ with four clear peaks at $n\pi/3$ ($n=0,1,2,3$).

\begin{figure}
\epsfig{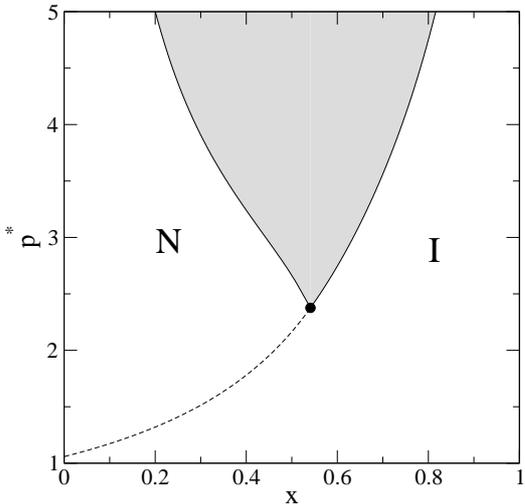}
\caption{Phase diagram of a binary mixture of the sublime triangle (species 1) and 
a triangle with $\kappa_2=5\kappa_1$, both having the same area.}
\label{fig9}
\end{figure}

To end this section we show in Fig. \ref{fig9} the usual I-N demixing scenario above a tricritical 
point occurring in binary mixtures of species with equal areas and sufficiently different aspect ratios.
The I-N and N-N demixing was also found in mixtures of two-dimensional particles 
with other shapes
(discorectangles, rectangles and ellipses \cite{yuri_demix,dani_demix,yuri_ellipses}) and they 
are well documented in mixtures of three-dimensional hard anisotropic bodies 
\cite{Vanakaras1,Camp,Vanakaras2,Varga4,Galindo}. 
We have chosen triangles with $\kappa_1=\sqrt{5+2\sqrt{5}}$ (the sublime triangle) 
and $\kappa_2=5\kappa_1$, and with the same particle areas. 
The second order I-N transition departs from the one-component fluid $x=0$ and extends up to a tricritical point, 
above which the mixture exhibits strong demixing. We should note that, at higher pressures, the I-N second order transition 
departing from the one-component fluid at $x=1$ should meet the I binodal
of the I-N demixing region at a critical end-point. The value of the corresponding pressure at 
this point is so high that our 
numerical scheme to find I-N coexistence, based on a Fourier expansion, becomes unstable.  

\section{Conclusions}
\label{conclusions}

We used SPT to study the phase behavior of the one-component HT fluid, focusing on 
a study of the relative stability and phase transitions between uniform phases. Isosceles  
HT exhibit a fascinating and rich phase diagram. We have shown that the TR phase exists  
not only for the equilateral triangles (with $\kappa=\sqrt{3}$), but also for a 
certain range of aspect ratios including this value. Also, we have found the presence of 
N-N transition ending in critical points for triangles with aspect ratios 
inside small intervals at both sides of $\sqrt{3}$. To our best knowledge this is the first example of 
one-component hard-particle systems exhibiting a N-N phase transition. We found that the 
I-N transition becomes of first order for aspect ratios less (greater) than the 
values $\kappa_1^{(\rm cep)}>\sqrt{3}$ ($\kappa_2^{(\rm cep)}<\sqrt{3}$), corresponding to 
two critical end-points where the I-N second-order line meets one of the binodals of the N-N transition at both 
sides of $\sqrt{3}$. For packing fractions below these values, the I-N transition 
is of second order, with the transition curve being asymmetric with respect to the change $\kappa \to\kappa^{-1}$: 
the obtuse triangles stabilize the N phase at lower packing fractions than the acute ones. The second order I-TR 
line intersects the I binodals of the first-order I-N transitions at two critical end-points 
that define the limits of the TR phase stability region, which in turn is bounded above  
by first-order TR-N transitions. 

Using the same formalism we have also studied some particular binary mixtures of HT. The 
main purpose was to elucidate if the mixing of triangles of certain geometries can 
stabilize a TR phase even when this phase is not stable in the two one-component limits. We have shown 
that if the species are symmetric with respect to $\sqrt{3}$, namely,  
(i) their aspect ratios fulfill the condition $\kappa_1\kappa_2=3$ and (ii) they have equal particle areas, 
there exists a range of aspect ratios, beyond that of the one-component fluid, for which 
a region of stable TR phase does exist. We computed the phase diagram (in the pressure-composition 
plane) for one of this particular mixtures, which exhibits a fascinating topology, i.e. the presence of first-order I-N and 
N-N transitions, the latter ending in a critical point. These transitions coalesce, at higher 
pressures, with the I-TR second-order transition line and further continue  
as two first-order TR-N transitions meeting at a Landau point. The TR binodals 
of both TR-N transitions are just the upper bounds of the region of 
TR stability. Finally, N binodals of a new N-N demixing transition depart from this Landau point,
with a demixing gap strongly increasing with pressure. This phase diagram resembles that of 
a binary symmetric mixture of rods and plates: the presence of a Landau point  
at the coalescence of both I-N transition curves, and N-N or biaxial N phase stability regions above, depending on 
the value of aspect ratios \cite{roij,yuri_prl}.

To end this section, we remark that the density expansion of the SPT is exact 
up to the second order, i.e. it recovers the exact second-virial coefficient, while it approximates 
the higher-order ones. In 2D the third- and higher-order virial coefficients, when properly scaled with the second
virial coefficient, do not vanish at the Onsager limit. Therefore, their effect 
on the phase behavior of 2D anisotropic particles is very important. 
In a previous study we have shown that the inclusion of the exact third-virial coefficient in a 
density-functional theory dramatically changes the location of the phase transitions between 
different orientationally ordered phases \cite{MR2}. The main effect of three-body correlations is
a substantial decrease in the packing fraction of the I-tetratic transition for all aspect ratios, which
entails an enlarged region of tetratic phase stability \cite{MR2}. 
The latter is an orientationally ordered phase with an orientational distribution function 
having fourfold symmetry: $f(\phi+n\pi/2)=f(\phi)$ ($n=1,\dots,4$). In an analogous way, we expect the
widening of the TR phase stability region in a HT fluid when the third virial coefficient is incorporated into the theory. 
In particular, the packing fraction of the I-TR transition is expected to decrease, a result supported by MC simulations 
conducted  on equilateral HT, where the transition was found at densities well below that estimated 
by SPT \cite{Dijkstra}. Recently a third virial theory of freely-rotating hard biaxial particles also resulted crucial to 
adequately predict the relative stability between orientationally ordered phases \cite{Marjorien}.

However, we believe the inclusion of non-uniform phases (such as the crystal phase)
would be an essential feature of more systematic studies in the future. In this line,
the recently derived fundamental-mixed-measure 
density-functional theory for hard freely-rotating 2D particles is at present the most promising theoretical tool to 
tackle this problem \cite{Wittmann}. 

\appendix
\section{Prove of the absence of a polar N phase}
\label{app1}
As pointed out already, the excluded area is minimal
when the main axes of the triangles are at a relative angle of $\pi$. This would in principle 
discard any polar nematic phase with
most of the particles pointing along a given direction. Instead, the HT will point with
equal probability along two equivalent directors differing in an angle of $\pi$. To prove
this result, suppose that the orientational distribution function $f(\phi;q)$ had the form
\begin{eqnarray}
	f(\phi;q)=(1-q)\tilde{f}(\phi)+q\tilde{f}(\pi-\phi), 
\end{eqnarray}
where $q$ is the probability that a given particle is oriented with respect to
the director pointing along the $\phi=\pi$ direction. Therefore, the particle will be oriented
with respect to a second director (pointing along the direction $\phi=0$) with probability $(1-q)$. The function
$\tilde{f}(\phi)$ is an orientational distribution function which in principle could have the property
$\tilde{f}(\phi)\neq \tilde{f}(\pi-\phi)$. Then the free-energy per particle 
(\ref{varphi}) is easily obtained as
\begin{eqnarray}
	&&\varphi=\log y -1 +\frac{1}{2\pi}\int_0^{2\pi} d\phi\Psi(\phi;q)\log \Psi(\phi;q)
	\nonumber\\
	&&+\frac{y}{2} \left[\sum_{j\geq 0} s_{2j}\left(\tilde{f}_{2j}\right)^2+
	(1-2q)^2\sum_{j\geq 0} s_{2j+1}
	\left(\tilde{f}_{2j+1}\right)^2\right].\nonumber\\ 
\end{eqnarray}
where $f(\phi;q)=(2\pi)^{-1}\Psi(\phi;q)$ while $\{\tilde{f}_n\}$ are the Fourier coefficients of 
$\tilde{f}(\phi)$. The derivative of $\varphi$ with respect to $q$ gives
\begin{eqnarray}
	S(q)&&\equiv \frac{\partial \varphi}{\partial q}=\frac{1}{2\pi}\int_0^{2\pi}d\phi 
	\Delta\tilde{\Psi}(\phi) 
	\log \Psi(\phi;q)\nonumber\\
	&&-2y(1-2q)\sum_{j\geq 1}s_{2j+1}\left(\tilde{f}_{2j+1}\right)^2,
\end{eqnarray}
where $\Delta\tilde{\Psi}(\phi)\equiv\tilde{\Psi}(\pi-\phi)-\tilde{\Psi}(\phi)$.
$S(q)$ is a monotonically increasing function because
\begin{eqnarray}
	S'(q)&&=\frac{1}{2\pi}\int_0^{2\pi}d\phi \frac{\left[\Delta\tilde{\Psi}(\phi)\right]^2}
	{\Psi(\phi;q)}
	\nonumber\\
	&&+4y\sum_{j\geq 0} s_{2j+1}\left(\tilde{f}_{2j+1}\right)^2\geq 0,
\end{eqnarray}
due to the positiveness of the odd coefficients: $s_{2j+1}>0$ $\forall j\geq 0$. Moreover,
using the symmetries $\Delta\tilde{\Psi}(\phi)=-\Delta\tilde{\Psi}(\pi-\phi)$
and $\Psi(\phi,\frac{1}{2})=\Psi(\pi-\phi;\frac{1}{2})$, it can be easily shown that $S(\frac{1}{2})=0$.
Thus the value $q=1/2$ is the only one that
satisfies the equilibrium condition $\displaystyle{\frac{\partial \varphi}{\partial q}=0}$.
Due to the distribution function symmetry,
$f(\phi,\frac{1}{2})=f(\pi-\phi,\frac{1}{2})$, the odd Fourier coefficients of the
expansion of $f(\phi,\frac{1}{2})$ are all equal to zero. We have shown numerically that 
this is the case by minimizing
the free-energy with respect to
all the Fourier coefficients, resulting in $\{f_{2j+1}=0\}$ $\forall \ j\geq 0$.
The main conclusion we can extract from the preceding analysis is that the main particle
axes are oriented with equal probability with respect to two equivalent, anti-parallel nematic directors.

\acknowledgments

Financial support under grants FIS2015-66523-P and FIS2013-47350-C5-1-R from Ministerio de 
Econom\'{\i}a, Industria y Competitividad (MINECO) of Spain is acknowledged.

\end{document}